\documentclass[showpacs,nofootinbib,preprintnumbers,prd,superscriptaddress,twocolumn,nofootinbib]{revtex4}

\usepackage{amsmath}
\usepackage{amsfonts}
\usepackage{amssymb}
\usepackage{graphicx}
\usepackage[titletoc]{appendix}
\usepackage{color}
\usepackage{hyperref}
\usepackage{cleveref}
\usepackage[rightcaption]{sidecap}
\usepackage{subfigure}
\usepackage{comment}

\usepackage{mathtools}

\usepackage{dcolumn}

\usepackage{array}
\usepackage{ctable}
\usepackage{multirow}
\usepackage{siunitx}
\usepackage{longtable}
\usepackage{tabularx}
\usepackage{booktabs}

\usepackage{amsmath}
\usepackage{amsfonts}
\usepackage{amssymb}
\usepackage[titletoc]{appendix}
\usepackage{hyperref}
\usepackage{cleveref}

\def\nat{Nature}

\def\prd{Phys. Rev. D}
\def\mnras{MNRAS}

\def\apj{ApJ}
\def\apjl{ApJ Lett.}
\def\apjs{ApJ Suppl. Ser.}

\def\aap{A\&A}

\def\jcap{JCAP}

\def\apss{Astrophysics and Space Science}

\begin{document}

\title{Model independent cosmographic constraints from DESI 2024}


\author{Orlando Luongo}
\email{orlando.luongo@unicam.it}
\affiliation{School of Science and Technology, University of Camerino, Via Madonna delle Carceri, Camerino, 62032, Italy.}
\affiliation{Istituto Nazionale di Fisica Nucleare (INFN), Sezione di Perugia, Perugia, 06123, Italy.}
\affiliation{SUNY Polytechnic Institute, 13502 Utica, New York, USA.}
\affiliation{INAF - Osservatorio Astronomico di Brera, Milano, Italy.}
\affiliation{Al-Farabi Kazakh National University, Al-Farabi av. 71, 050040 Almaty, Kazakhstan.}

\author{Marco Muccino}
\email{marco.muccino@lnf.infn.it}
\affiliation{School of Science and Technology, University of Camerino, Via Madonna delle Carceri, Camerino, 62032, Italy.}
\affiliation{Al-Farabi Kazakh National University, Al-Farabi av. 71, 050040 Almaty, Kazakhstan.}
\affiliation{ICRANet, P.zza della Repubblica 10, 65122 Pescara, Italy.}

\begin{abstract}
In this study, we explore model independent constraints on the universe kinematics up to the snap and jerk hierarchical terms. To do so, we consider the latest Baryon Acoustic Oscillation (BAO) release provided by the DESI collaboration, tackling the $r_d$ parameter to span within the range $[144,152]$ Mpc, with fixed step, $\delta r_d=2$ Mpc, aligning with Planck and DESI results. Thus, employing Monte Carlo Markov chain analyses, we place stringent constraints on the cosmographic series, incorporating three combinations of data catalogs: the first BAO with observational Hubble data, the second BAO with type Ia supernovae, and the last including all three data sets. Our results conclusively constrain the cosmographic series, say the deceleration $q_0$, the jerk $j_0$, and the snap $s_0$ parameters, at the $2$--$\sigma$ level, showcasing a significant departure on $j_0$ even at $1$--$\sigma$ confidence level, albeit being compatible with the $\Lambda$CDM paradigm on $q_0$ and $s_0$, at $2$--$\sigma$ level. Analogously, the $h_0$ tension appears  alleviated in the second hierarchy, say including snap. Finally, a direct comparison with the $\Lambda$CDM, $w$CDM models and the Chevallier-Polarski-Linder parametrization is reported, definitively favoring the wCDM scenario.

\end{abstract}



\maketitle

\section{Introduction}
The cosmological concordance model predicts the existence of a cosmological constant, $\Lambda$, entering the well-established  $\Lambda$CDM paradigm, quite robustly supported by several cosmological observations \cite{copeland}. However, the recent DESI’s map \cite{DESI} has analysed with higher precision than previous missions the underlying structure of the universe, showing results that, \emph{if confirmed},  may suggest a possible deviation to our comprehension toward dark energy. Accordingly, the recent experimental results made by Planck satellite \cite{Planck2018} and developments of dark energy evolution have been questioned.

Precisely, DESI measures the characteristic separation of galaxies, namely the baryonic acoustic oscillations (BAO). From these analyses, an apparent tension with the standard cosmological model arises up to $3.9$--$\sigma$ of confidence level, suggesting that a possible slightly evolving dark energy contribution, departing from the pure cosmological constant,  $\Lambda$ may exist\footnote{However, the DESI outcome represents another issue that increases the actual considerable conceptual and theoretical issues \cite{2022NewAR..9501659P} associated with the standard cosmological model, such as tensions \cite{DiValentino:2021izs}, fine-tuning and coincidence problem, cosmological constant problem \cite{LM} and so on \cite{revmia,mymodel}.}.

If, from the one hand, DESI seeks to answer whether dark energy has a constant value everywhere in the universe as a cosmological constant, from the other hand their results are based on assuming a given cosmological model \emph{a priori}, leading to a possible degeneracy problem. In other words, more than one model can explain dynamical dark energy.

Then, as we cannot be sure until we look at the next batch of data, albeit tempted to claim that the $\Lambda$CDM model is no longer suitable, it still appears as the best suite to describe the overall dynamics.

Indeed, due to the \emph{plethora} of contrasting  cosmological scenarios \cite{modelsDEbis,modelsDEtris}, the use of \emph{model independent treatments} to frame out the universe dynamics becomes decisive to single out the most suitable cosmological model \cite{modelsIND,modelsINDquinque}. Among all, a powerful example of model independent treatment is offered by \emph{cosmography} \cite{cosmography1,cosmography1bis,1974ApJ...192..577L,cosmography1tris}.

Essentially, Taylor series involving expansions of $a(t)$ and/or cosmic distances around our time \cite{cosmography2,cosmography2bis,cosmography2tris} are directly compared with cosmic data \cite{cosmography3}, without the need of postulating the cosmological model \emph{a priori} \cite{cosmography2,2023MNRAS.525.3104B,cosmography4,cosmography4tris,cosmography4quatris,miotr}. However, even if one can apply cosmography to different contexts, see e.g. Refs. \cite{2020EPJP..135....1C,2015PhRvD..91l4037C,2014PhRvD..90d4016C,2016JCAP...12..042D,2018JCAP...05..008C,2013PhRvD..87f4025A,2013PhRvD..87d4012A,2021MNRAS.503.4581L,2023MNRAS.518.2247L,2012MSAIS..19...37I}, it may produce non-conclusive numerical results due to intrinsic problems of convergence and truncation \cite{degeneracy1,degeneracy1bis,degeneracy1tris,degeneracy2,degeneracy2bis}. The need of \emph{further data points} is therefore essential in the realm of cosmographic expansions and, in this respect, \emph{the DESI measurements can significantly refine the cosmographic analysis} as we show throughout this work.

Motivated by the above points, we here analyse 30 model independent constraints of the universe kinematics. The first set of 15 bounds up to the jerk parameter, corresponding to the third expanded order of the luminosity distance, and the second set of 15 fits up to the further order, including the snap parameter. To do so, we perform a Monte Carlo Markov chain set of analyses, employing the most recent BAO measurements inferred by the DESI collaboration. Particularly, the $r_d$ parameter, associated with BAO data, is hereby taken in agreement with the result of the Planck satellite, also confirmed by DESI. Precisely, in our fits, we assume the sound horizon at the drag epoch, $r_d$, to span within the discrete interval $r_d\in[144,152]$ Mpc with a step of $\delta r_d=2$ Mpc. In such a way, we put stringent bounds on the cosmographic series, verifying the findings made by the DESI mission, in a model independent way. Accordingly, our findings are obtained by considering a hierarchy among cosmic data points. First, we consider the BAO points with the observational Hubble data (OHD), then we combine BAO points with the type Ia supernovae (SNe Ia) of the Pantheon catalog, and finally we use a combination of all the aforementioned three data sets. We report convergent limits on the cosmographic set, made up by the deceleration, $q_0$, the jerk $j_0$ and the snap $s_0$ parameters, showing that they turn out to be fully constrained at $1$--$\sigma$. We obtain very tight errors on the first hierarchy, up to the jerk parameter and, then, we compare our findings with dark energy models, among which the $\Lambda$CDM paradigm, the $w$CDM and the Chevallier-Polarski-Linder (CPL) parametrization. We conclude that dynamical dark energy better adapts to this catalog of BAO, since the concordance model exhibits severe departures mainly on the jerk parameter. Consequences on the onset of cosmic acceleration and on how to alleviate the cosmological tension on $h_0$ are thus discussed.

The paper is structured as follows. In Sect. \ref{warm}, we introduce the theoretical model independent expansions that we want to constrain. In Sect. \ref{numerical}, we report our numerical outcomes and analyse the corresponding bounds. The physical consequences of our analyses are also discussed. Conclusions and final outlooks are reported in Sect. \ref{conclusions}, together with perspectives of our work.

\begin{table*}
\centering
\setlength{\tabcolsep}{0.7em}
\renewcommand{\arraystretch}{1.4}
\begin{tabular}{lccccccc}
\hline\hline
Model           & $\Omega_m$ & $w$ or $w_0$ & $w_a$ & Bounds &
$q_0$                                   &
$j_0$                                &
$s_0$                                       \\
\hline
$\Lambda$CDM                            &
$0.310\pm0.007$ & $-1$ & $-$ &
Planck            & $-0.535\pm 0.011$ & 1 & $-0.395 \pm 0.033$\\
& $0.307\pm0.005$ & $-1$ & $-$ & DESI & $-0.540\pm0.008$& 1 & $-0.381\pm0.023$ \\
\hline
wCDM                            &
$0.310\pm0.007$ & $-0.960\pm0.080$ & $-$ &
Planck                                   & $-0.494\pm0.093$ & $0.881\pm0.230$ & $-0.564\pm0.230$\\
& $0.317\pm0.007$ & $-0.967\pm0.024$ & $-$ & DESI & $-0.491\pm0.034$ & $0.902\pm0.070$
 & $-0.566\pm
0.070$ \\
\hline
CPL                            &
$0.310\pm0.007$ & $-0.957\pm0.080 $ & $-0.290^{+0.320}_{-0.260}$ &
Planck    & $-0.490\pm0.093$ & $0.572^{+0.563}_{-0.501}$
 & $-1.806^{+1.421}_{-1.166}$\\
& $0.316\pm0.007$ & $-0.727\pm0.067$ & $-1.050^{+0.310}_{-0.270}$ & DESI & $-0.246\pm0.076$ & $-0.688^{+0.428}_{-0.387}$ & $-2.839^{+1.311}_{-1.237}$ \\
\hline
\end{tabular}
\caption{Values of the cosmographic coefficients inferred from the theoretical cosmographic sets for each model, namely Eqs. \eqref{lcdmset}, \eqref{wcdmset} and \eqref{cplset}. Here, we used the experimental values got from the Planck and DESI collaborations \cite{Planck2018,DESI}, as reported above. The corresponding error bars have been computed through a logarithmic error propagation. }
\label{tab:inferredvalues}
\end{table*}

\section{Theoretical warm up}\label{warm}

In a homogeneous, isotropic and spatially-flat universe, the corresponding Friedmann, Lema\^itre, Robertson and Walker line element
$ds^2=dt^2-a(t)^2\left[dr^2+r^2\,dl^2\right]$, with $dl^2\equiv (d\theta^2+\sin^2\theta\ d\phi^2)$, implies the presence of a single scale factor $a(t)$, conventionally normalized to the unity. With this metric, the Einstein equations provide the dynamical  Friedmann equations
\begin{equation}\label{eq:Friedmann}
H^2\equiv\left(\frac{\dot a}{a}\right)^2=\frac{8\pi G}{3}\rho\ ,\qquad \dot H+H^2 = -\frac{3P+\rho}{6}\,,
\end{equation}
where $H$ is the Hubble rate, while $\rho$ and $P$ are the energy density and pressure of the total cosmic fluid, respectively.

Once defining the cosmological densities $\rho_i$ as ratios $\Omega_i\equiv \rho_i/\rho_c$, normalized to the critical density $\rho_c\equiv 3H_0^2/(8\pi G)$, the cosmological content is well approximated by dust matter and dark energy, at late times, neglecting further fluids, such as radiation, neutrinos, etc.

The cosmological information is incorporated into the Hubble rate. Thus, to feature a model independent strategy to fix cosmic bounds, one can directly expand  the scale factor in Taylor series around the present time, with the purpose of fixing the derivatives of $a(t)$ at current time. This treatment is called cosmographic technique, namely an model independent investigation of the kinematic variables without taking into account any cosmological model \emph{a priori} \citep{2016IJGMM..1330002D}.

Cosmography becomes particularly interesting in view of the recent developments found by the DESI collaboration, where dynamical dark energy has not been fully-excluded and where an evident tension with previous studies arose \citep{DESI}.

Precisely, following the cosmographic recipe, we have
\begin{align}
\label{eq:scale factor jerk}
a^{(3)}\simeq &\,1 + H_0 (t-t_0) - {1\over2} q_0 H_0^2 (t-t_0)^2+\\
&\,{1\over3!} j_0 H_0^3 (t-t_0)^3 +{1\over4!} s_0 H_0^4 (t-t_0)^4+\ldots,\nonumber
\end{align}
where conventionally we truncated the series up to a given order, having
at all times \cite{auxiliarybis},
\begin{equation}
\label{toto}
q(t)\equiv -\dfrac{\ddot{a}}{aH^2}\quad,\quad j(t) \equiv \dfrac{\dddot{a}}{aH^3}\quad,\quad
s(t)\equiv \dfrac{\ddddot{a}}{aH^4}\,,
\end{equation}
as coefficients of the above expansion, in general known as \textit{deceleration} $q$, \textit{jerk} $j$, and \textit{snap} $s$ parameters.

Bearing the above considerations in mind and relating $a(t)$ to the redshift $z$, we then write the Hubble parameter as function of the luminosity distance $d_L(z)$ by
\begin{equation}
H(z)=\Big[\frac{d}{dz}\left(\frac{d_L(z)}{1+z}\right)\Big]^{-1}\,,
\end{equation}
immediately yielding
\begin{align}\label{dlandH}
d_L(z)=\frac{z}{H_0}\sum_{n=0}^{N}{\alpha_n\over n!}z^n\,,\quad
H(z)=\sum_{m=0}^{M}{H^{(m)}\over m!}z^m\,,
\end{align}
where the coefficients $\alpha_n$ and $H^{(m)}$ can be found in Refs.  \citep{revmia,Capozziello:2020ctn}.
The corresponding truncated series suffer from several problems, among which the order truncation and the convergence problem \cite{2016IJGMM..1330002D,gruber2014cosmographic,2012Ap&SS.342..155B}.

To possibly alleviating these issues, different hierarchies among data sets and orders of the employed series can be considered\footnote{We arbitrarily neglect in our analyses the possibility of working out auxiliary variables, as proposed in Refs. \cite{2012PhRvD..86l3516A,cosmography3,cosmography2,auxiliarybis} or alternatives to Taylor expansions as shown in Refs. \cite{gruber2014cosmographic,Capozziello:2020ctn}. }.
The orders and hierarchies that we here consider are 1) up to jerk parameter first and then, 2) up to snap parameter, all compared with three combinations of data sets, in the following order made by
{\small
\begin{itemize}
    \item[1.] BAO+OHD,
    \item[2.] BAO+SN,
    \item[3.] BAO+OHD+SN.
\end{itemize}
}
For our subsequent purposes, it is now convenient to compute the cosmographic series for three given models of dark energy.

Precisely, with $\Omega_m(z)\equiv \Omega_m(1+z)^3$, we focus on:

\begin{itemize}
    \item[-] {\bf The standard $\Lambda$CDM model}, with normalized Hubble rate $
E\equiv H/H_0=\sqrt{\Omega_{m}(z)+\Omega_\Lambda}$. The only model parameter is $\Omega_m$, being $\Omega_\Lambda=1-\Omega_m$, providing the current cosmographic set\footnote{The model fully degenerates with the dark fluid, that conversely to the $\Lambda$CDM scenario provides an evolving equation of state, albeit with constant pressure, see e.g. \cite{2014IJMPD..2350012L,Luongo:2012dv,2018PhRvD..98j3520L}.},
\begin{subequations}\label{lcdmset}
    \begin{align}
q_0&= \frac{3}{2}\Omega_m -1,\\
j_0&= 1,\\
s_0&= 1 - \frac{9}{2} \Omega_m.
    \end{align}
\end{subequations}
As an example that we will use later on, in a perfect concordance scenario, say $\Omega_m=0.3$, we obtain
\begin{align}\label{uno}
q_0=-0.55\quad,\quad j_0=1\quad,\quad s_0=-0.35\,.
\end{align}
    \item[-] {\bf The wCDM model}, in analogy to the $\Lambda$CDM model, has a negative barotropic factor likely due to a dynamical, slowly-rolling scalar field. The  model has two parameters, $\Omega_m$ and $w$, and the
    normalized Hubble rate is $
E=\sqrt{\Omega_{m}(z)+\Omega_X(1+z)^{3(1+ w )}}$, with  $\Omega_X=1-\Omega_m$ and
\begin{subequations}\label{wcdmset}
\begin{align}\label{q0l}
q_0= &\,\frac{1}{2} [1 - 3w ( \Omega_m - 1)],\\
j_0= &\,\frac{1}{2} [2-9 w (\Omega_m - 1) (1 + w )],\\
s_0 =&\, -{7\over2} + {9\over4} (\Omega_m-1) w [9 + w (16 + 9 w -\\
&\,3 \Omega_m (1 + w))],\nonumber
\end{align}
\end{subequations}
Again, in a genuine scenario, with arbitrary $\Omega_m=0.3$ and $w=-9/10$, we have:
\begin{align}\label{due}
q_0=-0.445,\quad j_0=0.717,\quad s_0=-0.706
\end{align}
\item[-] {\bf The CPL parametrization}, sometimes referred to as $w_0w_a$CDM model, originally found as a first order Taylor expansion around $a=1$ of the barotropic factor $
\omega(z)=\omega_0+\omega_a z/(1+z)$, suggests a normalized Hubble rate,
\begin{equation}\label{cpl2}
E=\sqrt{\Omega_{m}(z)+\Omega_{X}(1+z)^{3(1+\omega_0+\omega_a)}e^{-\frac{3\omega_a
z}{1+z}}}\,,
\end{equation}
where $\Omega_m$, $\omega_0$ and $\omega_a$ are the three, free model parameters, yielding a \emph{statistically more complicated model than the previous two}. We thus write
\begin{subequations}\label{cplset}
    \begin{align}\label{q0l}
q_0=&\,\frac{1}{2} [1 - 3\omega_0 (\Omega_m - 1)],\\
j_0=&\,\frac{1}{2} [2 - 9 \omega_0 (\Omega_m -1) (1 + \omega_0) +3 \omega_a -3 \Omega_m \omega_a ],\\
s_0=&\,{1\over4} \left\{-14 - 9 (-1 + \Omega_m) w_0 [-9 + w_0 (-16 - 9 w_0 +\nonumber\right.\\
&\,\left. 3 \Omega_m (1 + w_0))] -
   33 w_a + 33 \Omega_m w_a -\right.\nonumber\\
   &\,\left.9 (-7 + \Omega_m) (-1 + \Omega_m) w_0 w_a\right\}.
\end{align}
\end{subequations}

Analogously, with $\Omega_m=0.3$, $w_0=-9/10$ and $w_a=0.5$, we obtain
\begin{equation}\label{tre}
q_0=-0.445,\quad j_0=1.242,\quad s_0=1.155
\end{equation}
\end{itemize}
The values reported in Eqs. \eqref{uno}, \eqref{due} and \eqref{tre} are compatible with experimental results performed on them \cite{2016IJGMM..1330002D}. However, more precise values are found in Table \ref{tab:inferredvalues}, where we compute the cosmographic set through the bounds inferred from the Planck satellite and DESI measurements, respectively. Nevertheless, we now have all the ingredients to compare the truncated series in Eq. \eqref{eq:scale factor jerk}, directly with data and, then, to check the agreement of our results with DESI outcomes. Further, we will compare our expectations in Table \ref{tab:inferredvalues} with the cosmographic results later in the text.

\begin{table*}
\centering
\setlength{\tabcolsep}{1.7em}
\renewcommand{\arraystretch}{1.1}
\begin{tabular}{lcccc}
\hline\hline
Tracer  & $z_{\rm eff}$ & $d_M/r_d$         & $d_H/r_d$         & $d_V/r_d$     \\
\hline
BGS     & $0.30$        & $-$               & $-$               & $7.93\pm0.15$ \\
LRG     & $0.51$        & $13.62\pm0.25$    & $20.98\pm0.61$    & $-$           \\
LRG     & $0.71$        & $16.85\pm0.32$    & $20.08\pm0.60$    & $-$           \\
LRG+ELG & $0.93$        & $21.71\pm0.28$    & $17.88\pm0.35$    & $-$           \\
ELG     & $1.32$        & $27.79\pm0.69$    & $13.82\pm0.42$    & $-$           \\
QSO     & $1.49$        & $-$               & $-$               & $26.07\pm0.67$\\
Lya QSO & $2.33$        & $39.71\pm0.94$    & $8.52\pm0.17$     & $-$           \\
\hline
\end{tabular}
\caption{The DESI-BAO samples with tracers, effective redshifts $z_{\rm eff}$, and ratios $d_M/r_d$, $d_H/r_d$, and $d_V/r_d$. Reproduced from Ref. \cite{DESI}.}
\label{tab:BAO}
\end{table*}

\section{Numerical analysis}\label{numerical}

We combine the DESI-BAO measurements with standard low-redshift data surveys. The fixed catalog is thus the BAO measurements, to combined with all the other surveys.
Precisely, we consider the OHD \citep{2022LRR....25....6M} and the {\it Pantheon} catalog of SNe Ia \cite{2018ApJ...859..101S}.
Therefore, the general best-fit parameters are clearly determined directly by maximizing the total log-likelihood function
\begin{equation}
 \ln{\mathcal{L}} = \ln{\mathcal{L}_{\rm O}} + \ln{\mathcal{L}_{\rm S}} + \ln{\mathcal{L}_{\rm B}}\,.
\end{equation}
Clearly, the above relation represents the largest combinations of likelihoods, that reduces to the sum of two terms -- with BAO likelihood always present -- according to the kind of fit we perform.

Below, we define the contribution of each probe.
\begin{itemize}
\item[-] {\bf DESI-BAO data.} The DESI-BAO measurements are essentially galaxy surveys across $7$ distinct redshift bins within a total range, $z\in[0.1,4.2]$, in which all the measurements are effectively independent from each other, as remarked in Ref. \cite{DESI}. Accordingly, no covariance matrix is here introduced, whereas the systematics associated with these measurements introduce a negligible offset \cite{DESI,2021MNRAS.503.3510G}.

Hence, the DESI-BAO data are equivalently expressed as the ratios
\begin{equation}
\frac{d_M(z)}{r_d}=\frac{d_L(z)}{r_d(1+z)},\quad\frac{d_H(z)}{r_d}=\frac{c}{r_d H(z)}\,,
\end{equation}
or only by the ratio
\begin{equation}
\frac{d_V(z)}{r_d}=\frac{[z d_M^2(z)d_H(z)]^{1/3}}{r_d}\,,
\end{equation}
when the signal-to-noise ratio is low (see Table \ref{tab:BAO}).

The sound horizon at the drag epoch, $r_d$, depends upon the matter and baryon physical energy densities and the effective number of extra-relativistic degrees of freedom.
In our computation, we fit $r_d$ assuming different ranges of plausibility, i.e., enabling it to span within the values $r_d\in[144,152]$ Mpc, where both the Planck satellite and DESI-BAO expectations fall within. In so doing, we only \emph{a posteriori} extract viable values on it. The procedure of fixing $r_d$ is clearly motivated by reducing the complexity of computation. Even though a possible strategy could be that of letting $r_d$ be free to vary, the one above mentioned permits to significantly reduce the computations, without altering the final outputs.

Thus, with the assumption of Gaussian distributed errors, $\sigma_{X_i}$, the log-likelihood functions for each ratio, $X=d_M/r_d$, $d_H/r_d$, $d_V/r_d$, can be written as
\begin{equation}
\label{loglikeBAOu}
    \ln \mathcal{L}_{\rm X} = -\frac{1}{2} \sum_{i=1}^{N_{\rm X}}\left\{\left[\dfrac{X_i-X(z_i)}{\sigma_{X_i}}\right]^2 + \ln(2\pi\sigma^2_{X_j})\right\},
\end{equation}
and the total BAO log-likelihood turns out to be
\begin{equation}
\label{loglikeBAO}
    \ln \mathcal{L}_{\rm B} = \sum_X\ln \mathcal{L}_{\rm X}\,.
\end{equation}
\item[-] {\bf Hubble rate data.} We here adopt the most updated sample of OHD, consisting of $N_{\rm O}=34$ measurements, conventionally reported in Table~\ref{tab:OHD}. These measures are of particular interest since they are determined from spectroscopic detection of the differences in age, $\Delta t$, and redshift, $\Delta z$, of couples of passively evolving galaxies. To do so, we make the hypothesis that the underlying galaxies are however formed at the same time, enabling to consider the identity $H(z)=-(1+z)^{-1}\Delta z/\Delta t$ \citep{2002ApJ...573...37J}.

Consequently, the OHD systematics mostly depend upon stellar population synthesis models and libraries. Even the initial mass functions, taken into account with the purpose of calibrating the measures, together with the stellar metallicity of the population, may contribute, often with picks of further $20$--$30\%$ errors  \citep{2021MNRAS.501.3515M,2022LRR....25....6M,2023MNRAS.523.4938M}. The measures are thus not particularly accurate, albeit their determination is fully model independent.

For the sake of simplicity, again we employ  Gaussian distributed errors,  $\sigma_{H_k}$. So, the best-fit parameters are found by maximizing the log-likelihood
\begin{equation}
\label{loglikeOHD}
    \ln \mathcal{L}_{\rm O} = -\frac{1}{2} \sum_{i=1}^{N_{\rm O}}\left\{\left[\dfrac{H_i-H(z_i)}{\sigma_{H_i}}\right]^2 + \ln(2\pi\sigma^2_{H_i})\right\}\,.
\end{equation}

\begin{table}
\centering
\setlength{\tabcolsep}{1.5em}
\renewcommand{\arraystretch}{1.1}
\begin{tabular}{lcc}
   \hline\hline
    $z$     &$H(z)$ &  References \\
            &[km/s/Mpc]&\\
    \hline
    0.0708  & $69.0\pm 19.6\pm12.4^\star$ & \citep{Zhang2014} \\
    0.09    & $69.0 \pm12.0\pm11.4^\star$  & \citep{Jimenez2002} \\
    0.12    & $68.6\pm26.2\pm11.4^\star$  & \citep{Zhang2014} \\
    0.17    & $83.0\pm8.0\pm13.1^\star$   & \citep{Simon2005} \\
    0.1791   & $75.0  \pm 3.8\pm0.5^\dagger$   & \citep{Moresco2012} \\
    0.1993   & $75.0\pm4.9\pm0.6^\dagger$   & \citep{Moresco2012} \\
    0.20    & $72.9\pm29.6\pm11.5^\star$  & \citep{Zhang2014} \\
    0.27    & $77.0\pm14.0\pm12.1^\star$  & \citep{Simon2005} \\
    0.28    & $88.8\pm36.6\pm13.2^\star$  & \citep{Zhang2014} \\
    0.3519   & $83.0\pm13.0\pm4.8^\dagger$  & \citep{Moresco2016} \\
    0.3802  & $83.0\pm4.3\pm12.9^\dagger$  & \citep{Moresco2016} \\
    0.4     & $95.0\pm17.0\pm12.7^\star$  & \citep{Simon2005} \\
    0.4004  & $77.0\pm2.1\pm10.0^\dagger$  & \citep{Moresco2016} \\
    0.4247  & $87.1\pm2.4\pm11.0^\dagger$  & \citep{Moresco2016} \\
    0.4497  & $92.8 \pm4.5\pm 12.1^\dagger$  & \citep{Moresco2016} \\
    0.47    & $89.0\pm23.0\pm44.0^\dagger$     & \citep{2017MNRAS.467.3239R}\\
    0.4783  & $80.9\pm2.1\pm 8.8^\dagger$   & \citep{Moresco2016} \\
    0.48    & $97.0\pm62.0\pm12.7^\star$  & \citep{Stern2010} \\
    0.5929   & $104.0\pm11.6\pm4.5^\dagger$  & \citep{Moresco2012} \\
    0.6797    & $92.0\pm6.4\pm4.3^\dagger$   & \citep{Moresco2012} \\
    0.75    & $98.8\pm24.8\pm22.7^\dagger$     & \citep{2022ApJ...928L...4B}\\
    0.7812   & $105.0\pm9.4\pm6.1^\dagger$  & \citep{Moresco2012} \\
    0.80    & $113.1\pm15.1\pm20.2^\star$    & \citep{2023ApJS..265...48J}\\
    0.8754   & $125.0\pm15.3\pm6.0^\dagger$  & \citep{Moresco2012} \\
    0.88    & $90.0\pm40.0\pm10.1^\star$  & \citep{Stern2010} \\
    0.9     & $117.0\pm23.0\pm13.1^\star$  & \citep{Simon2005} \\
    1.037   & $154.0\pm13.6\pm14.9^\dagger$  & \citep{Moresco2012} \\
    1.26    & $135.0\pm60.0\pm27.0^\dagger$   & \citep{2023AA...679A..96T} \\
    1.3     & $168.0\pm17.0\pm14.0^\star$  & \citep{Simon2005} \\
    1.363   & $160.0 \pm 33.6^\ddag$  & \citep{Moresco2015} \\
    1.43    & $177.0\pm18.0\pm14.8^\star$  & \citep{Simon2005} \\
    1.53    & $140.0\pm14.0\pm11.7^\star$  & \citep{Simon2005} \\
    1.75    & $202.0\pm40.0\pm16.9^\star$  & \citep{Simon2005} \\
    1.965   & $186.5 \pm 50.4^\ddag$  & \citep{Moresco2015} \\
\hline
\end{tabular}
\caption{Updated OHD catalog with redshifts (first column), measurements with errors (second column), and references (third column). Systematic errors computed in this work are labeled with $\star$, with $\dagger$ if given by the literature, and with $\ddag$ when combined with statistical errors.}
\label{tab:OHD}
\end{table}
\item[-] {\bf SNe Ia.} The most recent catalog of SNe data points, namely the \emph{Pantheon} data set, is made up by $1048$ measures associated with SNe \cite{2018ApJ...859..101S}, albeit it can be significantly reduced, once the spatial curvature is assumed to be zero, to a catalog of $N_{\rm S}=6$ measurements, reported in a practical way as normalized Hubble rates,  $E_i$  \cite{2018ApJ...853..126R}.
Here, the corresponding log-likelihood function is given by
\begin{equation}
\label{loglikeSN}
\ln \mathcal{L}_{\rm S} = -\frac{1}{2}\sum_{i=1}^{N_{\rm S}} \left\{ \Delta\mathcal E_i^{\rm T} \mathbf{C}_{\rm S}^{-1}
\Delta\mathcal E_i + \ln \left(2 \pi |{\rm C}_{\rm S}| \right) \right\}\,,
\end{equation}
where we imposed $\Delta\mathcal E_i\equiv E_i^{-1} -  E^{-1}(z_i)$, the covariance matrix $\mathbf{C}_{\rm S}$ and its determinant $|{\rm C}_{\rm S}|$.
\end{itemize}

\subsection{Results}

We worked out two different hierarchies of coefficients, up to jerk first, and then snap. In both the above cited cases, the cosmographic series converge regardless the order and the value of $r_d\in[144,152]$ Mpc.

Concerning the bounds on $r_d$ and $H_0$, BAO measurements are sensitive to the degeneracy $H_0$--$r_d$ \cite{2012MNRAS.426.1280S}.
This is evident by noticing that in Tables \ref{tab:resultsj0}--\ref{tab:resultss0} the values of $H_0=100h_0$ km/s/Mpc decrease for increasing $r_d$, whereas all the other parameters are stable and insensitive to this degeneracy, regardless the hierarchy.
In this sense, it is within our expectations that DESI-BAO+SNe Ia fits -- with the SN data set based on $E(z)$ measurements, thus independent from $H_0$ -- are unable to break such a degeneracy.
As a result, for this set of fits, $r_d$ is absolutely unconstrained as well as $h_0$, that however aligns with Planck expectations \cite{Planck2018}.
\begin{table*}
\centering
\setlength{\tabcolsep}{0.9em}
\renewcommand{\arraystretch}{1.4}
\begin{tabular}{lcccccc}
\hline\hline
Data set                                &
$r_d$                                   &
$100h_0$                                &
$q_0$                                   &
$j_0$                                   &
$\ln \mathcal L$                        &
$\Delta$                               \\
& [Mpc] & [km/s/Mpc] & & & &           \\
\hline
DESI-BAO+OHD                            &
$144$                                   &
$70.24^{+1.32(+2.23)}_{-1.68(-2.56)}$   &
$-0.48^{+0.04(+0.07)}_{-0.05(-0.08)}$   &
$0.58^{+0.02(+0.04)}_{-0.01(-0.02)}$    &
$-141.64$                               &
$0.16$                                 \\
& $146$                                 &
$69.06^{+1.55(+2.44)}_{-1.53(-2.37)}$   &
$-0.47^{+0.04(+0.07)}_{-0.04(-0.07)}$   &
$0.58^{+0.02(+0.04)}_{-0.01(-0.02)}$    &
$-141.48$                               &
$0.00$                                 \\
& $148$                                 &
$68.21^{+1.47(+2.37)}_{-1.49(-2.41)}$   &
$-0.47^{+0.04(+0.07)}_{-0.05(-0.07)}$   &
$0.58^{+0.02(+0.04)}_{-0.01(-0.02)}$    &
$-141.61$                               &
$0.13$                                 \\
& $150$                                 &
$67.16^{+1.63(+2.53)}_{-1.30(-2.14)}$   &
$-0.48^{+0.05(+0.08)}_{-0.04(-0.07)}$   &
$0.58^{+0.02(+0.04)}_{-0.01(-0.02)}$    &
$-142.01$                               &
$0.53$                                 \\
& $152$                                 &
$66.59^{+1.31(+2.17)}_{-1.61(-2.41)}$   &
$-0.47^{+0.04(+0.07)}_{-0.05(-0.07)}$   &
$0.58^{+0.02(+0.04)}_{-0.01(-0.02)}$    &
$-142.66$                               &
$1.05$                                 \\
\hline
DESI-BAO+SNe                            &
$144$                                   &
$70.71^{+1.29(+2.08)}_{-1.38(-2.18)}$   &
$-0.50^{+0.04(+0.06)}_{-0.03(-0.06)}$   &
$0.58^{+0.01(+0.02)}_{-0.01(-0.02)}$    &
$14.88$                                 &
$0.00$                                 \\
& $146$                                 &
$69.74^{+1.27(+2.12)}_{-1.43(-2.25)}$   &
$-0.49^{+0.04(+0.06)}_{-0.04(-0.06)}$   &
$0.58^{+0.01(+0.02)}_{-0.01(-0.02)}$    &
$14.88$                                 &
$0.00$                                 \\
& $148$                                 &
$68.90^{+1.10(+1.97)}_{-1.46(-2.26)}$   &
$-0.49^{+0.03(+0.05)}_{-0.04(-0.06)}$   &
$0.58^{+0.01(+0.02)}_{-0.01(-0.02)}$    &
$14.88$                                 &
$0.00$                                 \\
& $150$                                 &
$67.92^{+1.24(+2.06)}_{-1.38(-2.17)}$   &
$-0.49^{+0.04(+0.06)}_{-0.04(-0.06)}$   &
$0.58^{+0.01(+0.02)}_{-0.01(-0.02)}$    &
$14.88$                                 &
$0.00$                                 \\
& $152$                                 &
$66.89^{+1.32(+2.15)}_{-1.26(-2.04)}$   &
$-0.49^{+0.03(+0.05)}_{-0.04(-0.07)}$   &
$0.58^{+0.01(+0.02)}_{-0.01(-0.02)}$    &
$14.88$                                 &
$0.00$                                 \\
\hline
DESI-BAO+OHD+SNe                        &
$144$                                   &
$70.17^{+1.24(+2.03)}_{-1.29(-2.01)}$   &
$-0.48^{+0.03(+0.05)}_{-0.04(-0.06)}$   &
$0.59^{+0.01(+0.02)}_{-0.01(-0.02)}$    &
$-125.16$                               &
$0.15$                                 \\
& $146$                                 &
$69.27^{+1.21(+2.00)}_{-1.31(-2.00)}$   &
$-0.48^{+0.03(+0.05)}_{-0.03(-0.06)}$   &
$0.59^{+0.01(+0.02)}_{-0.01(-0.02)}$    &
$-125.01$                               &
$0.00$                                 \\
& $148$                                 &
$68.41^{+1.23(+2.05)}_{-1.30(-2.08)}$   &
$-0.48^{+0.03(+0.05)}_{-0.04(-0.06)}$   &
$0.58^{+0.02(+0.03)}_{-0.01(-0.02)}$    &
$-125.15$                               &
$0.14$                                 \\
& $150$                                 &
$67.59^{+1.17(+1.92)}_{-1.28(-2.00)}$   &
$-0.48^{+0.03(+0.05)}_{-0.03(-0.05)}$   &
$0.59^{+0.01(+0.02)}_{-0.01(-0.02)}$    &
$-125.57$                               &
$0.56$                                 \\
& $152$                                 &
$66.69^{+1.23(+2.05)}_{-1.18(-1.90)}$   &
$-0.48^{+0.03(+0.05)}_{-0.04(-0.06)}$   &
$0.59^{+0.01(+0.02)}_{-0.01(-0.02)}$    &
$-126.24$                               &
$1.23$                                 \\
\hline
\end{tabular}
\caption{MCMC best-fit parameters and $1$--$\sigma$ ($2$--$\sigma$) error bars of the cosmographic series up to $j_0$.}
\label{tab:resultsj0}
\end{table*}
In addition, we report below the DESI-BAO and the OHD data points, in Tables \ref{tab:BAO}--\ref{tab:OHD}, respectively.

\begin{table*}
\centering
\setlength{\tabcolsep}{0.4em}
\renewcommand{\arraystretch}{1.4}
\begin{tabular}{lccccccc}
\hline\hline
Data set                                &
$r_d$                                   &
$100h_0$                                &
$q_0$                                   &
$j_0$                                   &
$s_0$                                   &
$\ln \mathcal L$                        &
$\Delta$                                \\
& [Mpc] & [km/s/Mpc] & & & & &          \\
\hline
DESI-BAO+OHD                            &
$144$                                   &
$70.15^{+2.62(+4.59)}_{-1.86(-3.48)}$   &
$-0.52^{+0.15(+0.23)}_{-0.09(-0.22)}$   &
$0.55^{+0.18(+0.53)}_{-0.11(-0.16)}$    &
$-0.47^{+0.26(+0.54)}_{-0.23(-0.39)}$   &
$-141.10$                               &
$0.66$                                 \\
& $146$                                 &
$68.70^{+2.59(+4.35)}_{-2.14(-3.65)}$   &
$-0.47^{+0.15(+0.22)}_{-0.11(-0.24)}$   &
$0.53^{+0.17(+0.47)}_{-0.09(-0.13)}$    &
$-0.46^{+0.18(+0.43)}_{-0.28(-0.45)}$   &
$-140.61$                               &
$0.17$                                 \\
& $148$                                 &
$68.12^{+2.36(+3.53)}_{-2.11(-3.59)}$   &
$-0.44^{+0.10(+0.19)}_{-0.16(-0.24)}$   &
$0.54^{+0.17(+0.33)}_{-0.10(-0.13)}$    &
$-0.48^{+0.24(+0.34)}_{-0.27(-0.44)}$   &
$-140.44$                               &
$0.00$                                 \\
& $150$                                 &
$67.25^{+2.51(+3.59)}_{-2.37(-3.69)}$    &
$-0.42^{+0.10(+0.17)}_{-0.20(-0.26)}$   &
$0.53^{+0.22(+0.37)}_{-0.10(-0.13)}$    &
$-0.43^{+0.17(+0.29)}_{-0.32(-0.51)}$   &
$-140.56$                               &
$0.12$                                 \\
& $152$                                 &
$65.92^{+2.25(+3.78)}_{-2.54(-3.80)}$   &
$-0.40^{+0.13(+0.21)}_{-0.17(-0.23)}$   &
$0.52^{+0.15(+0.26)}_{-0.08(-0.13)}$    &
$-0.52^{+0.17(+0.33)}_{-0.31(-0.49)}$   &
$-140.80$                               &
$0.36$                                 \\
\hline
DESI-BAO+SNe                            &
$144$                                   &
$71.01^{+1.08(+1.99)}_{-1.87(-2.56)}$   &
$-0.49^{+0.07(+0.10)}_{-0.07(-0.12)}$   &
$0.57^{+0.13(+0.24)}_{-0.09(-0.16)}$    &
$-0.46^{+0.18(+0.32)}_{-0.22(-0.38)}$   &
$14.89$                                 &
$0.02$                                 \\
& $146$                                 &
$69.65^{+1.35(+2.38)}_{-1.51(-2.28)}$   &
$-0.49^{+0.05(+0.09)}_{-0.09(-0.13)}$   &
$0.59^{+0.16(+0.22)}_{-0.10(-0.15)}$    &
$-0.46^{+0.14(+0.27)}_{-0.23(-0.34)}$   &
$14.88$                                 &
$0.01$                                 \\
& $148$                                 &
$69.84^{+1.43(+2.22)}_{-1.39(-2.15)}$   &
$-0.48^{+0.05(+0.09)}_{-0.09(-0.12)}$   &
$0.58^{+0.13(+0.21)}_{-0.10(-0.15)}$    &
$-0.47^{+0.20(+0.30)}_{-0.22(-0.38)}$   &
$14.87$                                 &
$0.00$                                 \\
& $150$                                 &
$67.94^{+1.30(+2.11)}_{-1.31(-2.13)}$   &
$-0.51^{+0.08(+0.11)}_{-0.07(-0.10)}$   &
$0.60^{+0.14(+0.20)}_{-0.11(-0.18)}$    &
$-0.46^{+0.16(+0.31)}_{-0.23(-0.36)}$   &
$14.90$                                 &
$0.03$                                 \\
& $152$                                 &
$67.07^{+1.42(+2.12)}_{-1.40(-2.33)}$   &
$-0.50^{+0.07(+0.11)}_{-0.07(-0.12)}$   &
$0.61^{+0.12(+0.22)}_{-0.12(-0.18)}$    &
$-0.46^{+0.18(+0.30)}_{-0.23(-0.38)}$   &
$14.89$                                 &
$0.03$                                 \\
\hline
DESI-BAO+OHD+SNe                        &
$144$                                   &
$70.53^{+1.29(+2.17)}_{-1.38(-2.08)}$   &
$-0.50^{+0.07(+0.11)}_{-0.06(-0.09)}$   &
$0.57^{+0.14(+0.21)}_{-0.08(-0.13)}$    &
$-0.50^{+0.17(+0.26)}_{-0.19(-0.28)}$   &
$-124.74$                               &
$0.50$                                 \\
& $146$                                 &
$69.49^{+1.27(+2.01)}_{-1.43(-2.28)}$   &
$-0.50^{+0.06(+0.11)}_{-0.05(-0.10)}$   &
$0.59^{+0.10(+0.18)}_{-0.09(-0.14)}$    &
$-0.51^{+0.16(+0.28)}_{-0.21(-0.33)}$   &
$-124.30$                               &
$0.06$                                 \\
& $148$                                 &
$68.69^{+1.38(+2.17)}_{-1.45(-2.32)}$   &
$-0.50^{+0.06(+0.14)}_{-0.08(-0.12)}$   &
$0.57^{+0.14(+0.21)}_{-0.07(-0.15)}$    &
$-0.49^{+0.14(+0.24)}_{-0.21(-0.33)}$   &
$-124.24$                               &
$0.00$                                 \\
& $150$                                 &
$67.75^{+1.32(+2.33)}_{-1.46(-2.35)}$   &
$-0.49^{+0.07(+0.11)}_{-0.06(0.11)}$    &
$0.59^{+0.12(+0.20)}_{-0.09(-0.15)}$    &
$-0.50^{+0.15(+0.25)}_{-0.21(-0.34)}$   &
$-124.41$                               &
$0.17$                                 \\
& $152$                                 &
$67.09^{+1.10(+1.99)}_{-1.48(-2.28)}$   &
$-0.49^{+0.06(+0.09)}_{-0.07(-0.11)}$   &
$0.61^{+0.11(+0.18)}_{-0.10(-0.15)}$    &
$-0.55^{+0.18(+0.28)}_{-0.18(-0.28)}$   &
$-124.84$                               &
$0.60$                                 \\
\hline
\end{tabular}
\caption{MCMC best-fit parameters and $1$--$\sigma$ ($2$--$\sigma$) error bars of the cosmographic series up to $s_0$.}
\label{tab:resultss0}
\end{table*}

In general, the best-fit parameters, including the comoving sound horizons at the drag epoc $r_d$, have been determined by looking at the set of model parameters furnishing the maximum value of the log-likelihood $\ln \mathcal L$, i.e., when $\Delta=0.00$.
Excluding the DESI-BAO+SNe fits, in BAO+OHD and BAO+OHD+SNe Ia combinations of data sets, the minimum $\Delta$ is related to $r_d\simeq146$ Mpc for the hierarchy up to $j_0$, and to $r_d\simeq148$ Mpc for the hierarchy up to $s_0$.
The corresponding contour plots are portrayed in Figs. \ref{fig:fits1}--\ref{fig:fits3}.

Comparing the outcomes on the parameters $q_0$, $j_0$ and $s_0$ with previous cosmographic bounds, it is evident that this is among the first occurrences in which the cosmographic series can always converge.
From all the measurements, it appears evident that $q_0<0, s_0<0$, while $j_0>0$, as theoretically predicted in Ref. \cite{Luongo:2013rba}. However, quite surprisingly, the values of the parameters are stable regardless the hierarchial order involved, namely in both the cases and for all the data sets involved.

In particular, from the sign of $q$, one can infer whether (and how much) the universe is decelerating, while a positive sign of $j_0$ indicates a transition time between the matter to dark energy dominated phases.

This point appears crucial, if one combines $j$ with the variation of $q$. Indeed, one can write,
\begin{equation}
q^\prime\equiv\frac{dq}{dz}=\frac{j-2q^2-q}{1+z}\,,
\end{equation}
again in a model independent way. So, at $z=0$, we infer that  the onset of acceleration deviates from the standard models presented in Eqs. \eqref{uno}, \eqref{due} and \eqref{tre}, say
\begin{subequations}
    \begin{align}
q^\prime_{\rm _{\Lambda CDM}}&=0.95\ \ ,\ \ q^\prime_{\rm _{wCDM}}=0.77\ \ ,\ \ q^\prime_{\rm_{CPL}}=1.29\,,\\
\label{thiswork}
q^\prime_{\rm_{cosmo}}&=0.58\,,
    \end{align}
\end{subequations}
where for the cosmographic approach in Eq. \eqref{thiswork} we used $q_0=-0.5$ and $j_0=0.58$. The $q^\prime$ predicted from the model independent measurements from cosmography is incredibly in line with the wCDM model and departs from the $\Lambda$CDM and the CPL models.

The above conclusion directly follows from all the results on $j_0$, independently from the hierarchy, that strongly departs from the standard cosmological model even at $1$--$\sigma$. This is much more evident, up to $2$--$\sigma$ for the first hierarchy, up to $j_0$. The error bars are extremely small in this case and are in line with $j_0\simeq 0.6$.
These values of $j_0$ are however clearly in tension with Eq. \eqref{uno}, while compatible with Eq. \eqref{due}, specially for the hierarchy up to $s_0$. However, there is an evident tension with the values of $j_0$ determined in the CPL parametrization, Eq. \eqref{tre}.

Unlike previous expectations, we do not find agreement with models of dark energy that predict $j_0\geq1$, representing however the standard case found previously in the literature.

Conversely, within the hierarchy up to $s_0$, the results on $q_0$ and $s_0$ match the standard model predictions within $1$--$\sigma$. The situation is different for the results of the hierarchy up to $j_0$, that leave open the chance to recover the standard model expectations at $2$--$\sigma$.

From the above results, DESI-BAO measurements deviate severely the final outputs, namely they are responsible to deviate from the pure cosmological constant dark energy.

As a final double check, we can analyse the results in Table \ref{tab:inferredvalues} with our model independent results from Tables \ref{tab:resultsj0} and \ref{tab:resultss0}. We find good agreement on $j_0$ and $s_0$ for dynamical dark energy models with respect to our cosmographic constraints. We find a severe tension on $j_0$ with respect to the concordance paradigm, the $\Lambda$CDM model. However, we find a departing outcome on $q_0$ for the CPL model. Consequently, we conclude that our predictions align with cosmographic results mainly for dynamical dark energy, especially for the wCDM model. However, we cannot conclude univocally that the $\Lambda$CDM model appears disfavored and, then, the need of further data release will clarify the numerical tensions on $j_0$ and $q_0$ inferred from our analyses.

In addition, a very relevant fact is related to the $H_0$ tension, occurring between the two values,
\begin{subequations}
    \begin{align}
    h_0^P&=0.674\pm0.005,\quad {\rm Planck\,\,measurement,}\\
    h_0^R&=0.730 \pm 0.010,\quad {\rm Riess\,\,measurement.}
    \end{align}
\end{subequations}
since in our cases, the Hubble tension seems to be alleviated, within the $2$--$\sigma$ measurements mainly as the snap parameter is plugged into the analysis. The same does not occur as the hierarchy is up to jerk. Once again, even though this \emph{is not} an evidence for that dark energy can alleviate the tension itself, can instead suggest that a dynamical fluid driving the universe to speed up can in principle naturally fix the strong discrepancy between local observations and Planck measurements. This appears quite surprisingly, as previous BAO data points indicated the opposite evidence, see e.g. \cite{2023PDU....4201298A}.

\section{Discussion}\label{conclusions}

The recent DESI measurements have opened new avenues toward the physics of dark energy. Even though there is no consensus toward an evolving dark energy fluid, the corresponding BAO appear in tension with the standard cosmological model. The results by the DESI collaboration \cite{DESI} are based on the assumptions that an underlying cosmological model is fixed \emph{a priori}.

In this work, we considered a model independent treatment to bound cosmic data points, using the BAO catalog from the DESI mission. To do so, we here found model independent constraints of the universe kinematics, with two main hierarchies, the first up to the jerk term, namely up to the third order of the luminosity distance expansion, and then up to the snap term, corresponding to the fourth order.

In so doing, we worked out a Monte Carlo Markov chain set of analyses, based on three combinations of data sets: the first making use of BAO and OHD data, the second replacing OHD with SNe Ia, while the last combining all three data sets. Particularly, following the expectations got from the Planck satellite, the $r_d$ parameter, associated with BAO measurements, was here assumed to span within the interval $r_d=144$--$152$ Mpc, over a finite grid of values with step $\delta r_d=2$ Mpc.

We thus found more stringent bounds on the cosmographic series. First, for both the hierarchies the error bars were particularly small. The cosmographic series, conversely to previous cases, converged and provided results that departed significantly from the standard cosmological model, especially for what concerns $j_0$ that appears strongly in tension with the fixed value predicted by the $\Lambda$CDM model, namely $j_0=1$.

On the other hand, our findings certified that $q_0$ and $s_0$ could agree with the standard concordance scenario, even at the level of $1$--$\sigma$ for each fit.

We also compared the inferred cosmographic series, Table \ref{tab:inferredvalues}, with the numerical one that we obtained from our fits, in Tables \ref{tab:resultsj0}-\ref{tab:resultss0}. We found a better agreement on $j_0$ and $s_0$ using the wCDM model, while the CPL parametrization appears not fully-constrained, especially at the level of $q_0$. However, the concordance model appears favored to constrain $s_0$, albeit huge discrepancies are associated with $j_0$. Clearly, for all the above cases, we showed that there is the chance to fix the free parameters of given dark energy models, in order to set $q_0,j_0$ and $s_0$ to agree with DESI data.

Hence, from our analyses, it is then particularly evident that the tension on $j_0$ cannot be avoided within the $\Lambda$CDM model. This implies that the variation of $q$ appears quite different than the one predicted by the standard cosmological model, resorting the idea that dark energy contribution may evolve in time.  Remarkably, we showed that the cosmological tensions on $h_0$ appears alleviated especially in the second hierarchy that considers Taylor series up to the snap term.

Concluding, even if we found a better matching between our cosmographic outcomes and dynamical dark energy models, the CPL parametrization may unexpectedly suffer from the tension on $q_0$, whereas wCDM appears favored than CPL.

As perspective, it would be crucial to wait for future mission data release that will be given in the near future. So, our results can be refined in view of additional points. As above stated, moreover, the covariance matrix has not been considered in our analysis, following the recipe suggested in the original DESI paper \cite{DESI}, where it seems not to be needful. Hence, working it out with the new data release will be crucial to disclose the goodness of this new catalog. Meanwhile, it will be interesting to clarify the role of spatial curvature in model independent fits, as well as possible alternatives to the standard Taylor series, such as Pad\'e expansions \cite{pade1,pade2} or Chebychev series \cite{chebyschev} to find refined cosmic constraints. In analogy to these proposals, we intend to work cosmography out by investigating the role of auxiliary variables and to perform possible high-redshift cosmokinematics \cite{2012PhRvD..86l3516A}, adapting different series to the incoming sets of data points.

\section*{Acknowledgements}
OL expresses his gratitude to Alejandro Aviles for private discussions related to the subject of this work. The work by OL is partially financed by the Ministry of Education and Science of the Republic of Kazakhstan, Grant: IRN AP19680128. The work by MM is partially financed by the Ministry of Education and Science of the Republic of Kazakhstan, Grant: IRN BR21881941.

\appendix

\section*{Contour plots}\label{contourplots}

We here display the contour plots of the best fit models of our analyses in the following figures.

\begin{figure*}
\centering
{\includegraphics[width=0.42\hsize,clip]{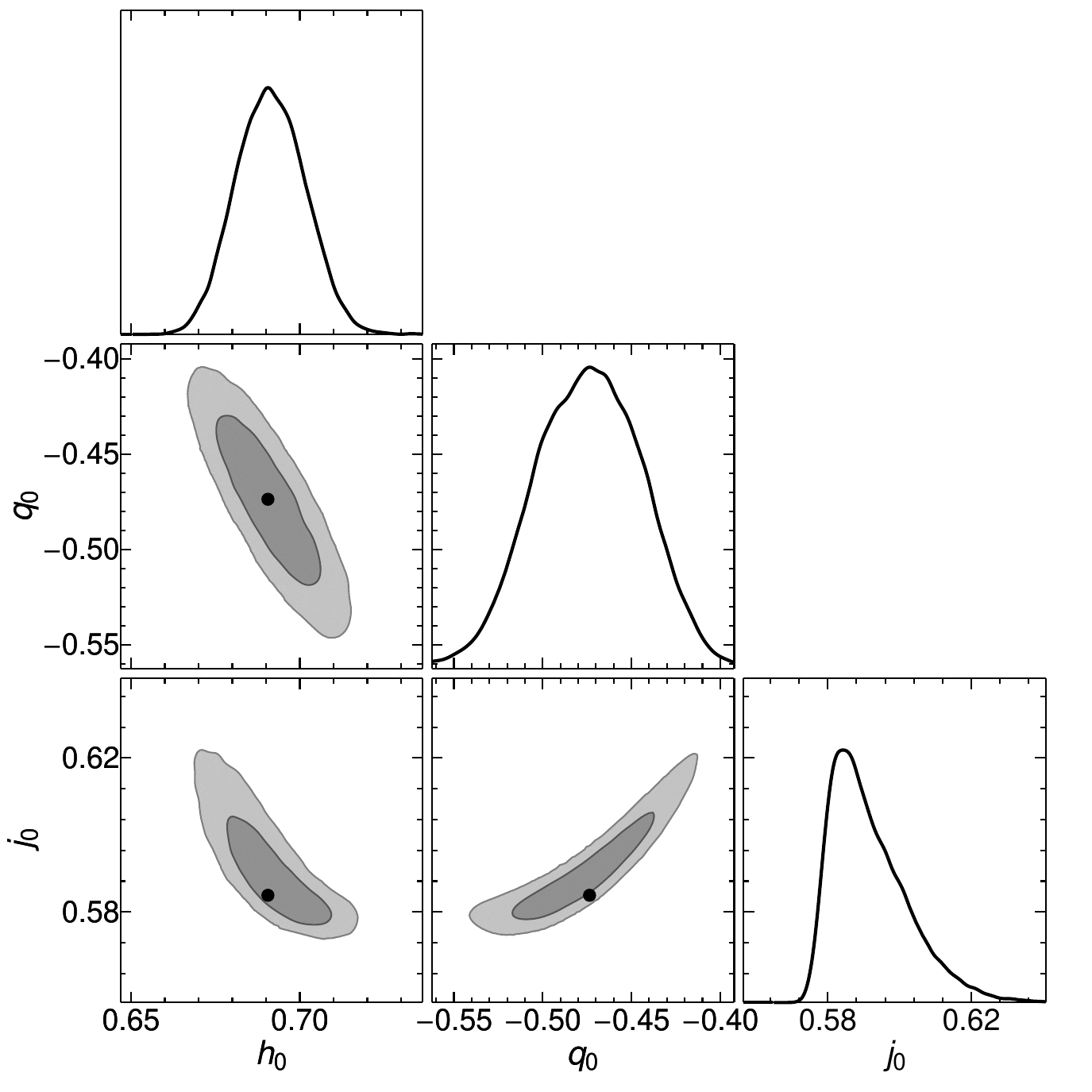}}
\hfill
\includegraphics[width=0.56\hsize,clip]{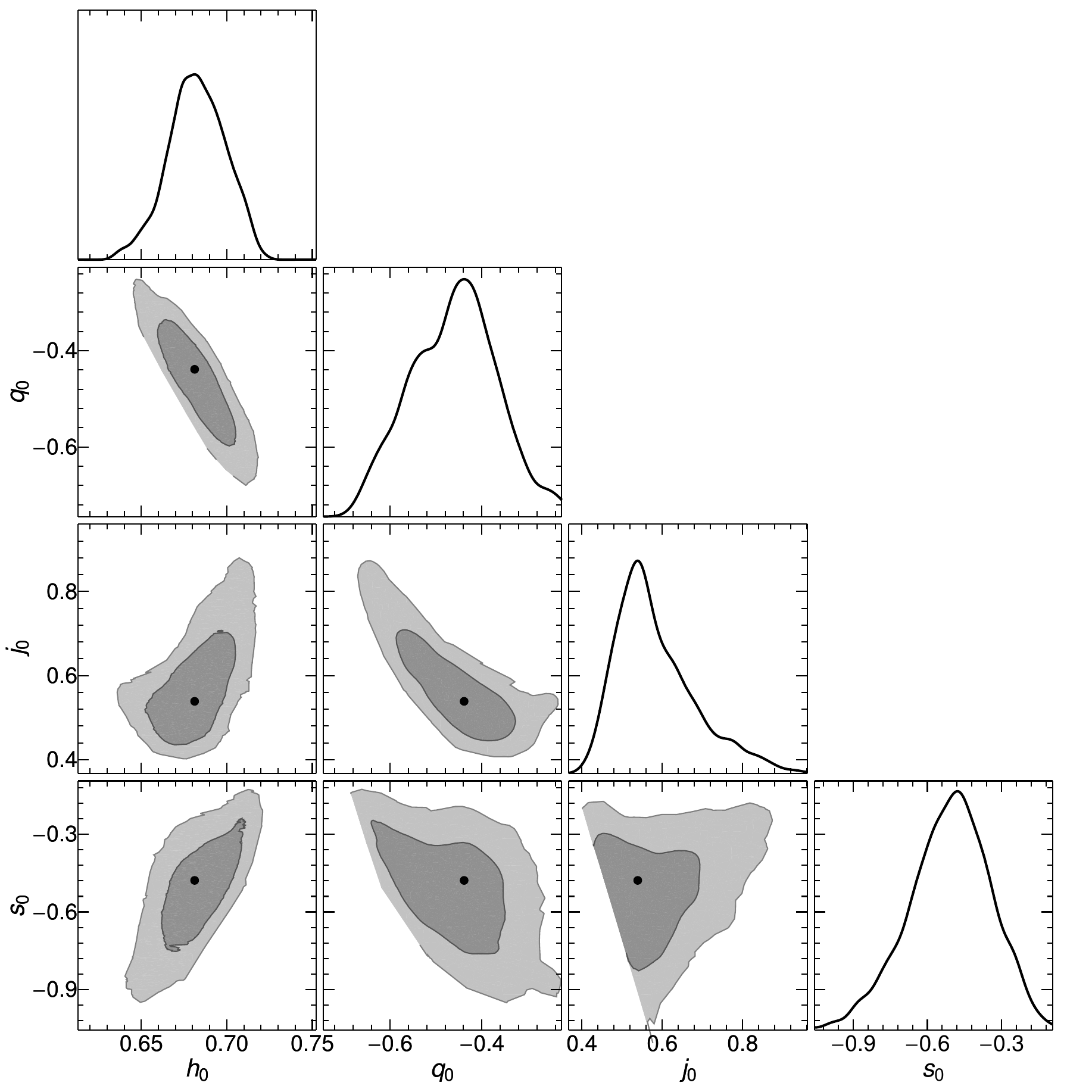}
\caption{DESI-BAO+OHD contour plots of the cosmographic series up to $j_0$ (left) and up to $s_0$ (right). The best-fit parameters are indicated by black circles, whereas the $1$-$\sigma$ ($2$-$\sigma$) contours are shown as dark (light) gray areas.}
\label{fig:fits1}
\end{figure*}

\begin{figure*}
\centering
{\includegraphics[width=0.42\hsize,clip]{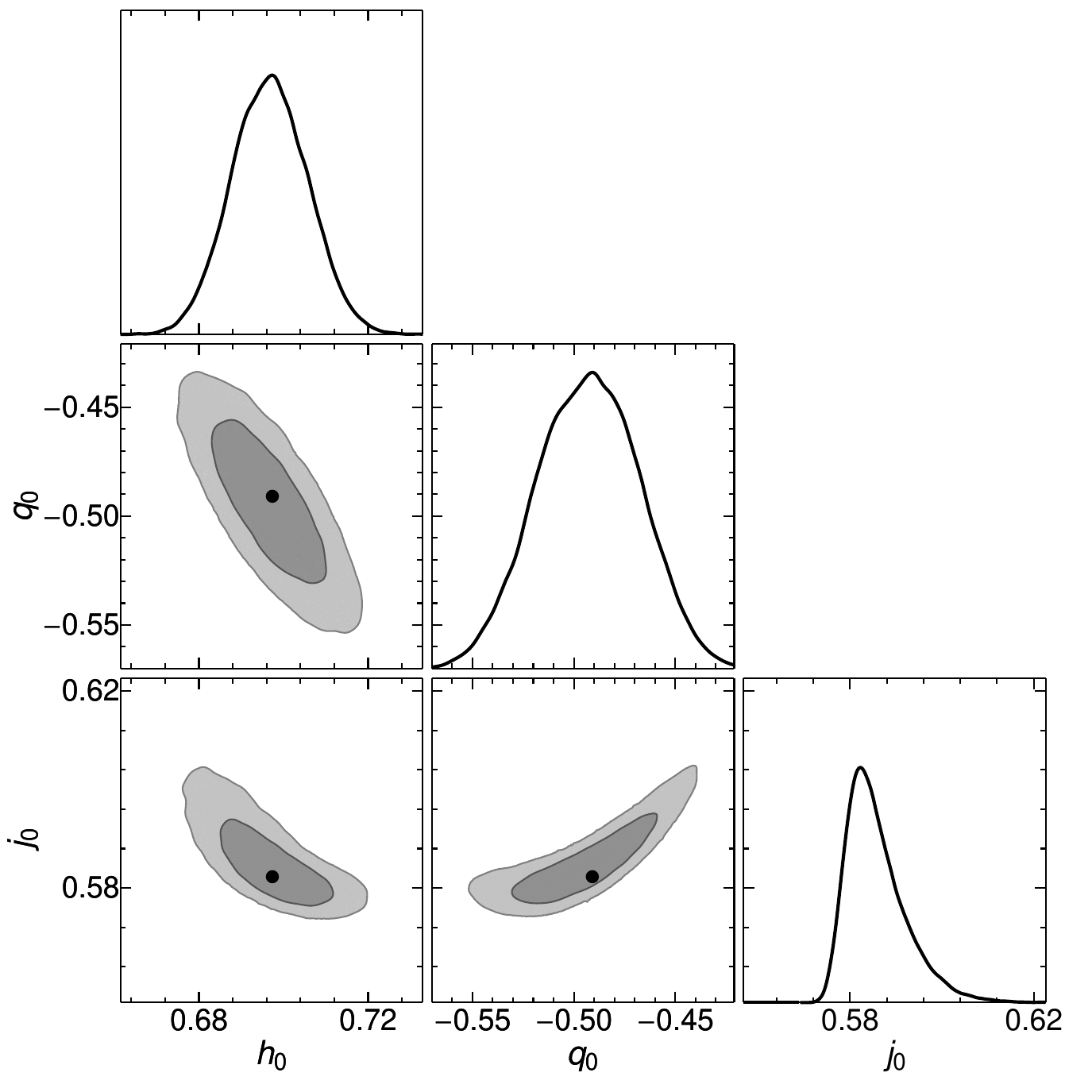}
\hfill
\includegraphics[width=0.56\hsize,clip]{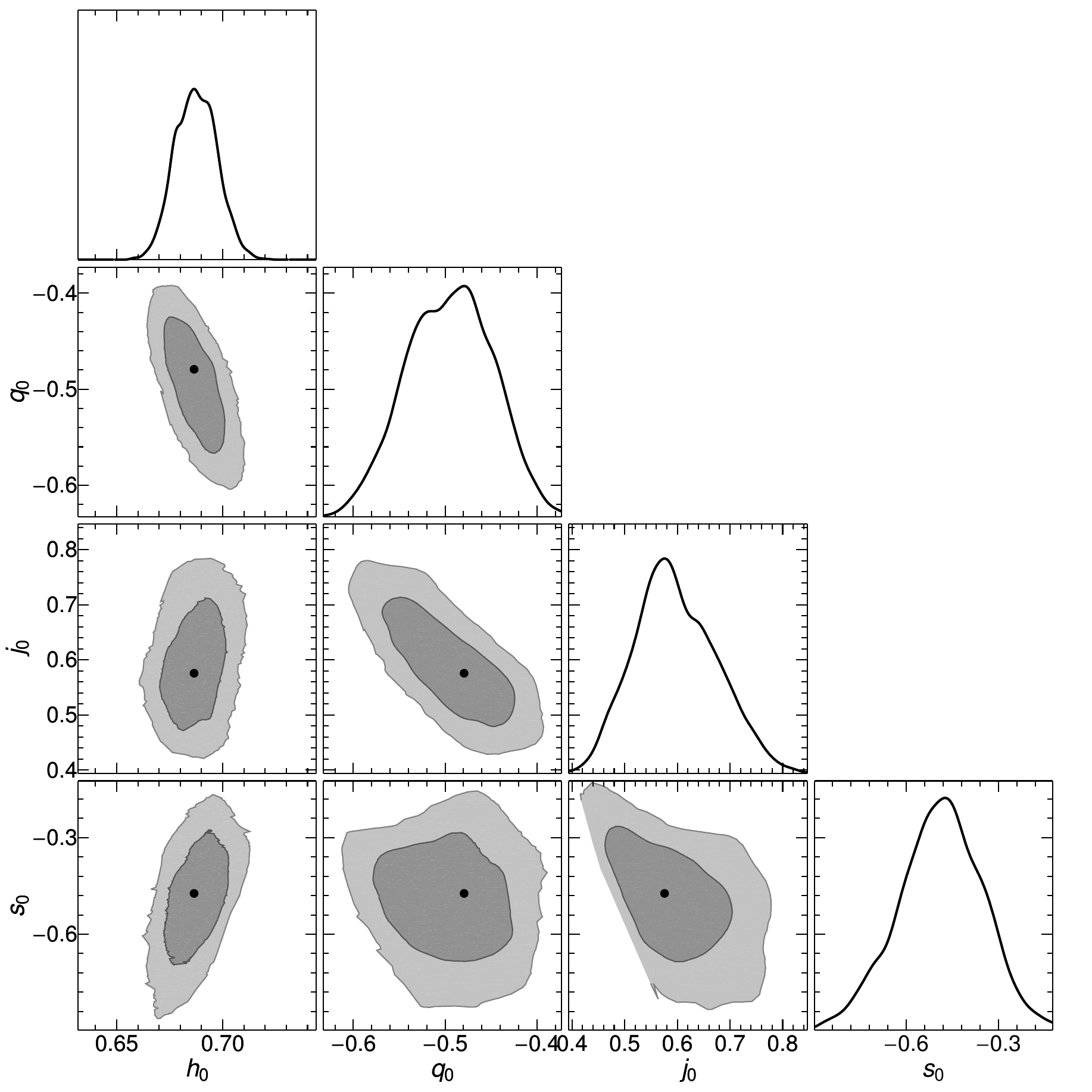}}
\caption{DESI-BAO+SNe contour plots of the cosmographic series up to $j_0$ (left) and up to $s_0$ (right). Symbols and colors have the same meaning as in Fig. \ref{fig:fits1}}
\label{fig:fits2}
\end{figure*}

\begin{figure*}
\centering
{\includegraphics[width=0.42\hsize,clip]{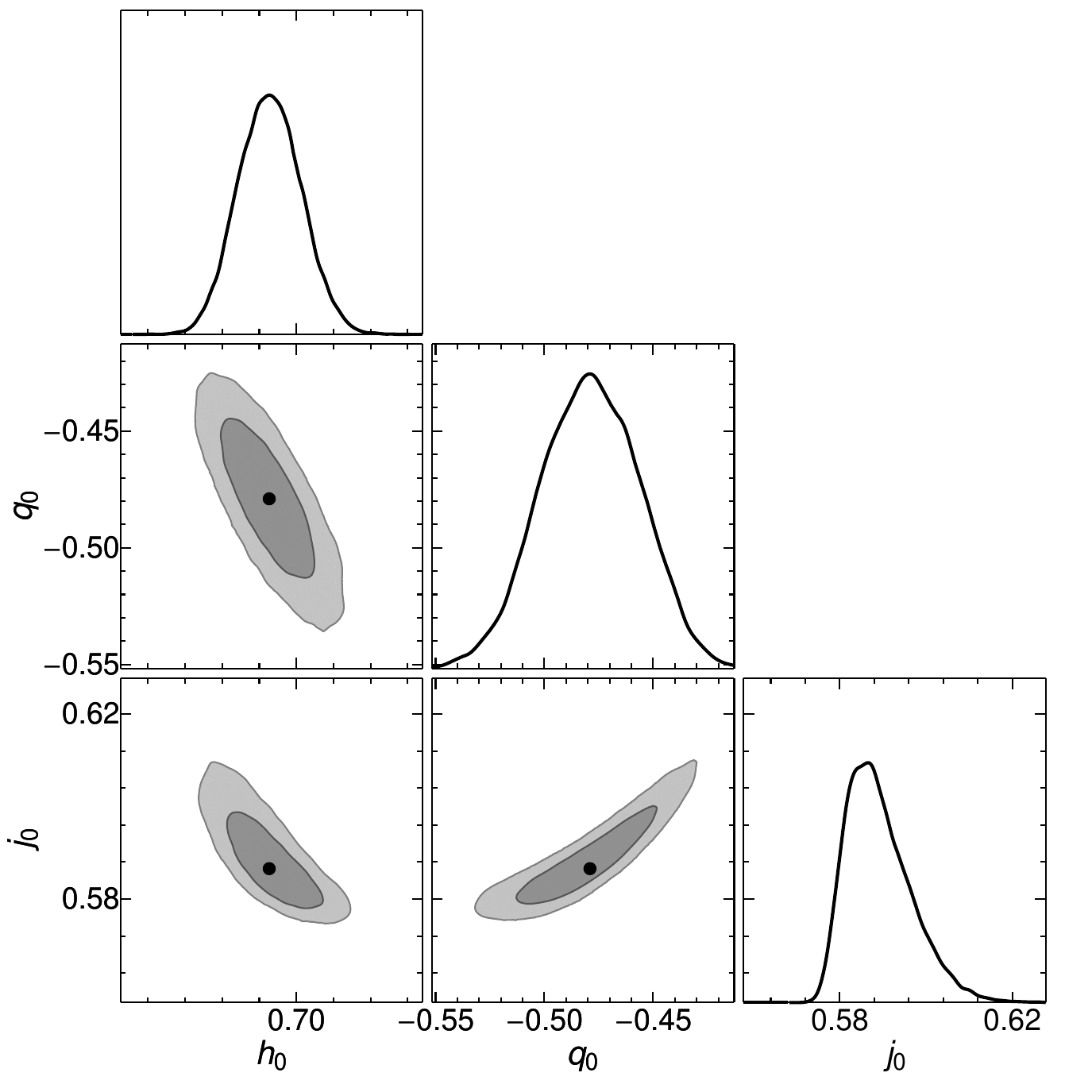}
\hfill
\includegraphics[width=0.56\hsize,clip]{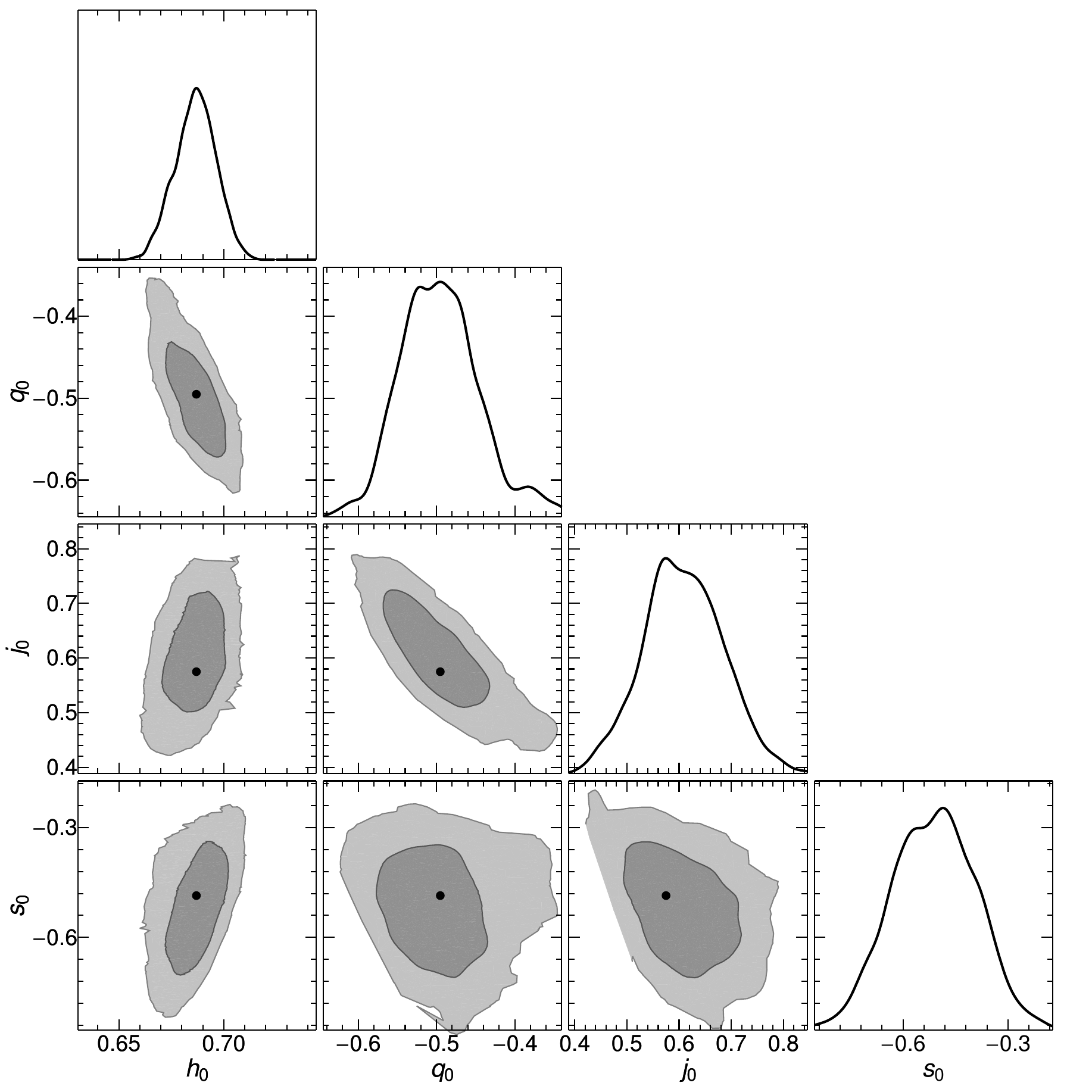}}
\caption{DESI-BAO+OHD+SNe contour plots of the cosmographic series up to $j_0$ (left) and up to $s_0$ (right). Symbols and colors have the same meaning as in Fig. \ref{fig:fits1}.}
\label{fig:fits3}
\end{figure*}

As it appears evident from Figs. \ref{fig:fits1}--\ref{fig:fits3} above, the fits align with the final outcomes predicted by the DESI collaboration, albeit the jerk parameter favors values smaller than unity, as expected by previous efforts \cite{Capozziello:2019cav}.

\end{document}